\pdfoutput=1
\documentclass[aps,prc,twocolumn,floatfix,superscriptaddress,nofootinbib,preprintnumbers,longbibliography,amsmath,amsthm,amssymb]{revtex4-2}
\usepackage{graphicx}

\usepackage{amsfonts}
\usepackage{color}
\usepackage[colorlinks=true,linkcolor=blue,citecolor=blue]{hyperref}
\usepackage{dcolumn}
\usepackage{bm}
\usepackage{braket}

\newcommand{\be}{\begin{equation}}
\newcommand{\ee}{\end{equation}}

\newcommand{\br}{\boldsymbol{r}}

\newcommand{\ii}{\mathrm{i}}

\begin{document}
\allowdisplaybreaks[1]

\title{Enhanced moments of inertia for rotation in neutron-rich nuclei
}

\author{Kenichi Yoshida}
\email[E-mail: ]{kyoshida@ruby.scphys.kyoto-u.ac.jp}
\affiliation{Department of Physics, Kyoto University, Kyoto, 606-8502, Japan}

\preprint{KUNS-2924}
\date{\today}

\begin{abstract}
Ground-state moments of inertia (MoI) are investigated 
for about 1700 even-even nuclei 
from the proton drip line to the neutron drip line 
up to $Z=120$ and $N=184$.
The cranked Skyrme--Hartree--Fock--Bogoliubov 
equation is solved in the coordinate space.
This model describes well the available experimental 
data of more than 300 nuclides possessing an appreciable deformation. 
I find that the MoI greatly increase near the drip line 
whereas the deformation is not as strong as 
estimated by the empirical relation. 
Systematic measurements of the excitation energy  
and the transition probability to the first $I^\pi=2^+$ 
state in neutron-rich nuclei not only 
reveal the evolution of deformation 
but can also constrain an effective pair interaction.
\end{abstract}

\maketitle

\section{Introduction}\label{intro}
Nuclear rotational motion emerges due to the spontaneous breaking of the rotational symmetry~\cite{BM2}. 
As stepping away from the magic number, 
the first $I^\pi=2^+$ state becomes lower in energy: 
The collective mode of excitation changes its character 
from the vibration to the rotation as the deformation develops.
A na\"ive question arises here: How strong should the deformation be for the picture of the rotation to be well-drawn?

Recently, various spectroscopic studies have been carried out 
to explore unique structures in neutron-rich nuclei. 
The excitation energy of the $2_1^+$ state, 
$E(2^+_1)$, is often among the first quantities accessible in experiments 
and systematic measurements have revealed the 
evolution of the shell structure~\cite{SEASTAR,gad15,ots20}.
Besides the change of the shell structure associated with the onset of deformation, 
the $E(2_1^+)$ value may provide rich information on exotic nuclei.  
A significant lowering of $E(2^+_1)$ observed in 
a near-drip-line nucleus $^{40}$Mg 
could be a signal of new physics in drip-line nuclei~\cite{cra19}, 
as the theoretical calculations 
have predicted that the magnitude of 
deformation is not enhanced in $^{40}$Mg
comparing with the Mg isotopes with less neutrons~\cite{ter97,rod02,yos09a,yos10,yao11,wat14,rod16,shi16}.

The pair correlation is present in the ground state 
and plays a decisive role in describing various phenomena 
such as the energy gap in spectra of even-even nuclei, 
the odd-even staggering in binding energies and so on~\cite{rin80,bri05}.
Furthermore, the pairing is indispensable for 
a strong collectivity of the low-frequency quadrupole 
vibration~\cite{mat12}
and a reduced value of moments of inertia (MoI) for rotation 
from the rigid-body estimation~\cite{rin80}.
Therefore, the $E(2_1^+)$ value should be 
scrutinized by taking not only 
the deformation but the superfluidity into account. 

Another critical issue in exploring the drip-line nuclei 
is a need of the careful treatment of the asymptotic 
part of the nucleonic density.
An appropriate framework is Hartree--Fock--Bogoliubov (HFB) theory, 
solved in the coordinate-space representation~\cite{bul80,dob84}.
This method has been used extensively 
in the description of spherical systems but is much more difficult to implement for systems with deformed equilibrium shapes. 
Therefore, calculations have been 
mostly restricted to axially symmetric nuclei~\cite{ter03,bla05,yos08a,oba08,pei08,kas21}. 
A standard technique 
to describe the non-axial shapes is to employ 
a truncated single-particle basis, 
which consists of localized states and discretized-continuum 
oscillating states, 
for solutions of the HFB equation~\cite{ter96}. 
Such a method should not be able to describe adequately 
the spatial profile of densities at large distances. 
Recently, 
the HFB equation has been solved 
by employing the contour integral technique 
and the shifted Krylov subspace method for the 
Green's function~\cite{jin17,kas20} to circumvent the 
successive diagonalization of the matrix with huge dimension.

In this work, 
I investigate the rotational motion 
in neutron-rich nuclei near the drip line 
with emphasizing the pairing.
At high spins where the pairing vanishes, 
proposed is a novel mechanism of a nucleus being bound 
beyond the neutron drip line~\cite{afa19}. 
Here, I study the lowest spin state, namely the $2^+_1$ state, 
in even-even non-spherical nuclei. 
A key quantity is the MoI for rotation: 
$E(2_1^+)=6/2\mathcal{J}$. 

\section{Model and method}

The MoI is evaluated microscopically by the Thouless--Valatin 
procedure or the self-consistent cranking model as 
$\mathcal{J}=\lim_{\omega_{\rm rot} \to 0}\frac{J_x}{\omega_{\rm rot}}$
with $J_x=\langle \hat{J}_x \rangle$ 
and $\omega_{\rm rot}$ being the rotational frequency about the $x$-axis~\cite{rin80}.
I solve the cranked HFB (CHFB) equation to obtain the MoI and 
take the natural units: $\hbar =c=1$. 

The numerical procedure to solve 
the CHFB equation is described in Ref.~\cite{yos22}:
I impose the reflection symmetry 
about the $(x, y)$-, $(y,z)$- and $(z, x)$-planes. 
Thus, the parity $\mathfrak{p}_k$ $(=\pm 1)$ 
and $x$-signature $r_k$ ($=\pm \ii$) are a good quantum number. 
I solve the CHFB equation by diagonalizing the HFB Hamiltonian 
in the three-dimensional (3D) Cartesian-mesh 
representation with the box boundary condition. 
Thanks to the reflection symmetries, 
I have only to consider the octant region explicitly in space 
with $x\ge0$, $y\ge0$, and $z\ge0$; see Refs.~\cite{bon87,oga09} for details.
I use a 3D lattice mesh $x_i=ih-h/2, y_j=jh-h/2, z_k=kh-h/2 \ \  (i,j,k=1,2,\cdots M)$ with a mesh size $h=1.0$ fm and $M=12$ for each direction. 
A reasonable convergence with respect to the mesh size $h$ 
and the box size $M$ is 
obtained for not only drip-line nuclei 
but medium-mass nuclei~\cite{yos22}.
For diagonalizing the HFB matrix, 
I use the ScaLAPACK {\sc pdsyev} subroutine~\cite{ScaLAPACK}. 
A modified Broyden's method~\cite{bar08} is utilized to calculate new densities during the self-consistent iteration. 
The quasiparticle energy is cut off at 60 MeV.
For each iteration, 
it takes about 10 core hours at the 
SQUID computer facility (composed of 1520 nodes of Intel Xeon Platinum 8368 processor) 
of Osaka University. 
To obtain the convergence, typically, 50--100 iterations are needed. 
Therefore, the total cost for a systematic calculation of 
1700 nuclei is about 17000 node hours.

\section{Results and discussion}
\subsection{Validity of the present model}

The MoI of the ground state is evaluated 
at $\omega_{\rm rot}=0.05$ MeV.
I employed the SkM*~\cite{bar82} and SLy4~\cite{cha98} functionals 
augmented by  
the Yamagami--Shimizu--Nakatsukasa (YSN) pairing-density functional~\cite{yam09}, 
which is given as
\begin{equation}
    \mathcal{E}_{\rm pair}(\br)=\dfrac{V_0}{4}
    \sum_{\tau={\rm n},{\rm p}}g_\tau[\rho,\rho_1]|\tilde{\rho}_\tau(\br)|^2
    \label{pair_DF}
\end{equation}
with
\begin{equation}
    g_\tau[\rho,\rho_1]=1-\eta_0 \dfrac{\rho(\br)}{\rho_0}
    -\eta_1\dfrac{\tau_3 \rho_1(\br)}{\rho_0}
    -\eta_2\left[\dfrac{\rho_1(\br)}{\rho_0} \right]^2.
    \label{YSN_pairing}
\end{equation}
Here, $\rho(\br)$ and $\rho_1(\br)$ are the isoscalar and isovector densities,  
$\tau={\rm n}$ (neutron) or ${\rm p}$ (proton), and $\rho_0=0.16$ fm$^{-3}$ 
is the saturation density of symmetric nuclear matter. 
The parameters $V_0, \eta_0, \eta_1, \eta_2$ 
were optimized to reproduce the experimental pairing gaps globally 
and are summarized in Table III of Ref.~\cite{yam09}. 
Note that the parameters for the $\rho_1$ dependence $\eta_1, \eta_2$ are positive.
The YSN pairing functional was constructed based on the finding 
that the inclusion of the isospin dependence in the pairing
functional gives a good reproduction of the pairing gaps 
in both stable and neutron-rich nuclei 
and in both symmetric nuclear matter and in neutron matter~\cite{mar07,mar08}.

\begin{figure*}[t]
\begin{center}
\includegraphics[scale=0.85]{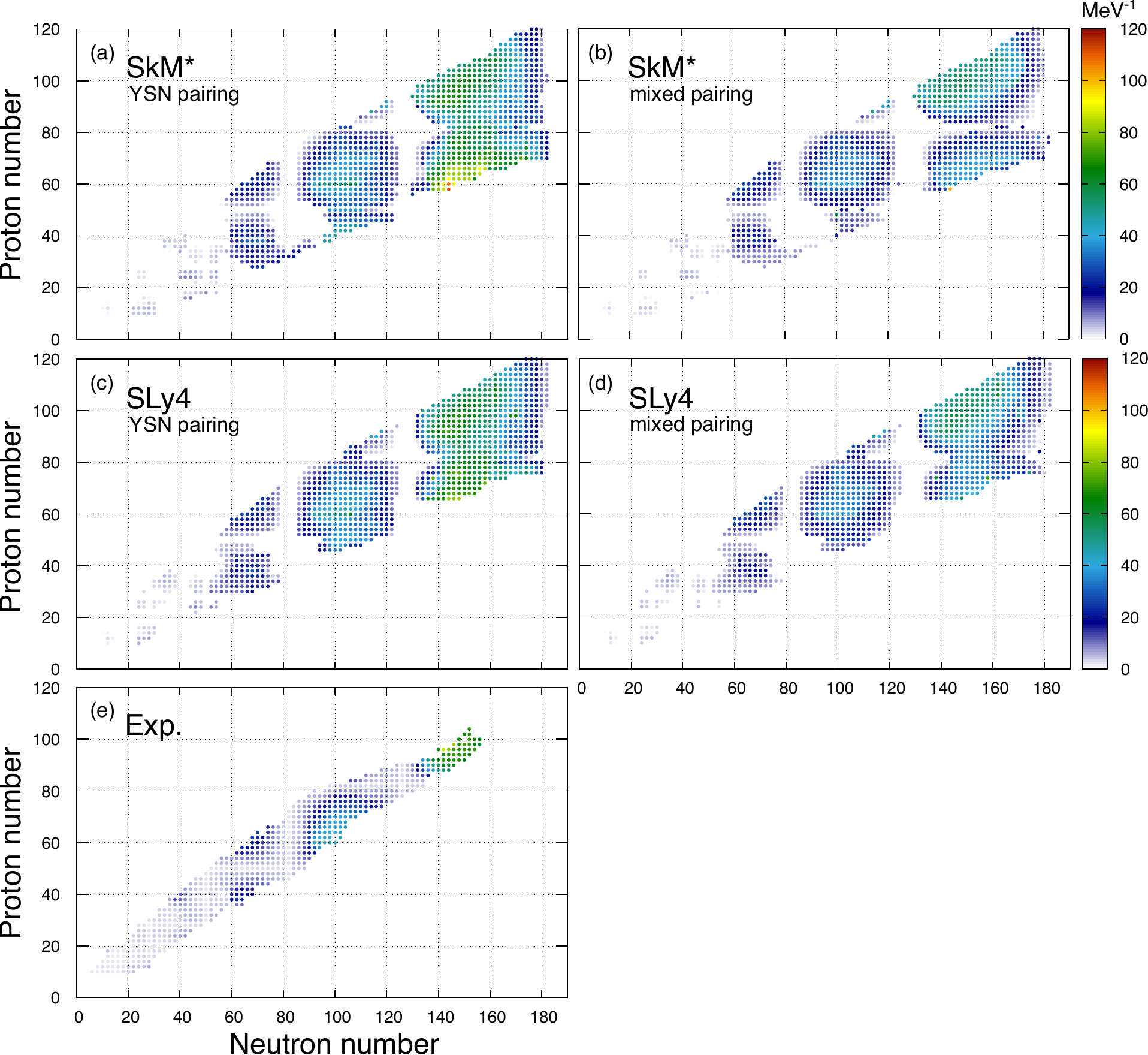}
\caption{\label{fig:def_MoI} 
Calculated moments of inertia $\mathcal{J}$ for the SkM* and SLy4 functionals. 
The experimental data are taken from Ref.~\cite{NNDC}, which is evaluated as $3/E(2^+_1)$. 
}
\end{center}
\end{figure*}

There are 657 even-even nuclei with known $E(2^+_1)$~\cite{NNDC}.
In the present study here I limit the scope by excluding the very light nuclei ($Z<10$), for which mean-field theory is least justified. 
This eliminates 22 nuclei. 
The experimental data  
evaluated as $3/E(2^+_1)$ for 635 nuclei are displayed in
Fig.~\ref{fig:def_MoI}(e).
There is no collective rotation in spherical nuclei 
where the MoI is zero. 
Actually, I defined the spherical nuclei if the calculated 
MoI is less than 0.1 MeV$^{-1}$. 
An additional 273 (260) nuclei 
have been eliminated for that reason, leaving 362 (375) nuclei in the present analysis.

\begin{figure}[t]
\begin{center}
\includegraphics[scale=0.24]{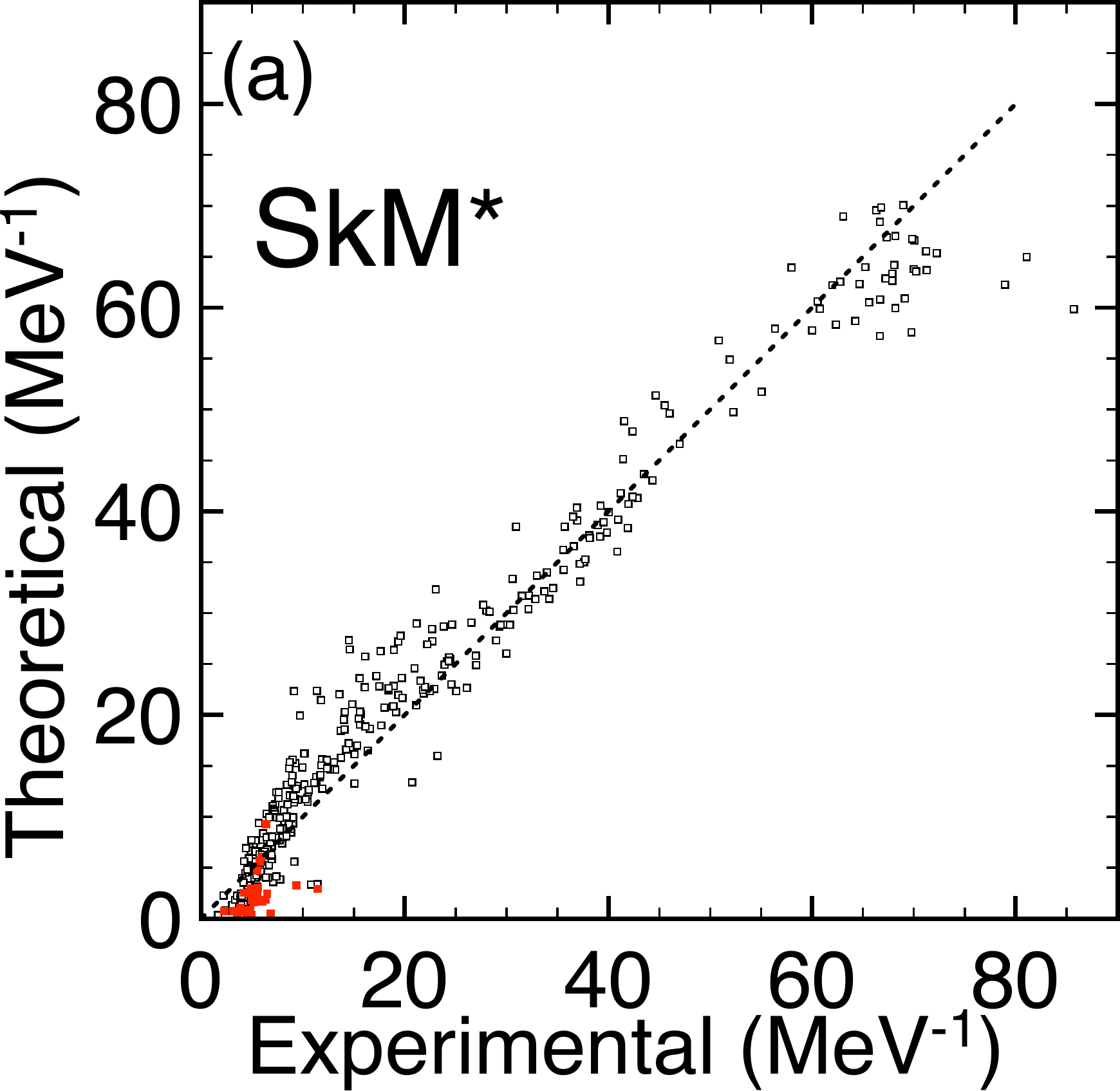}
\includegraphics[scale=0.24]{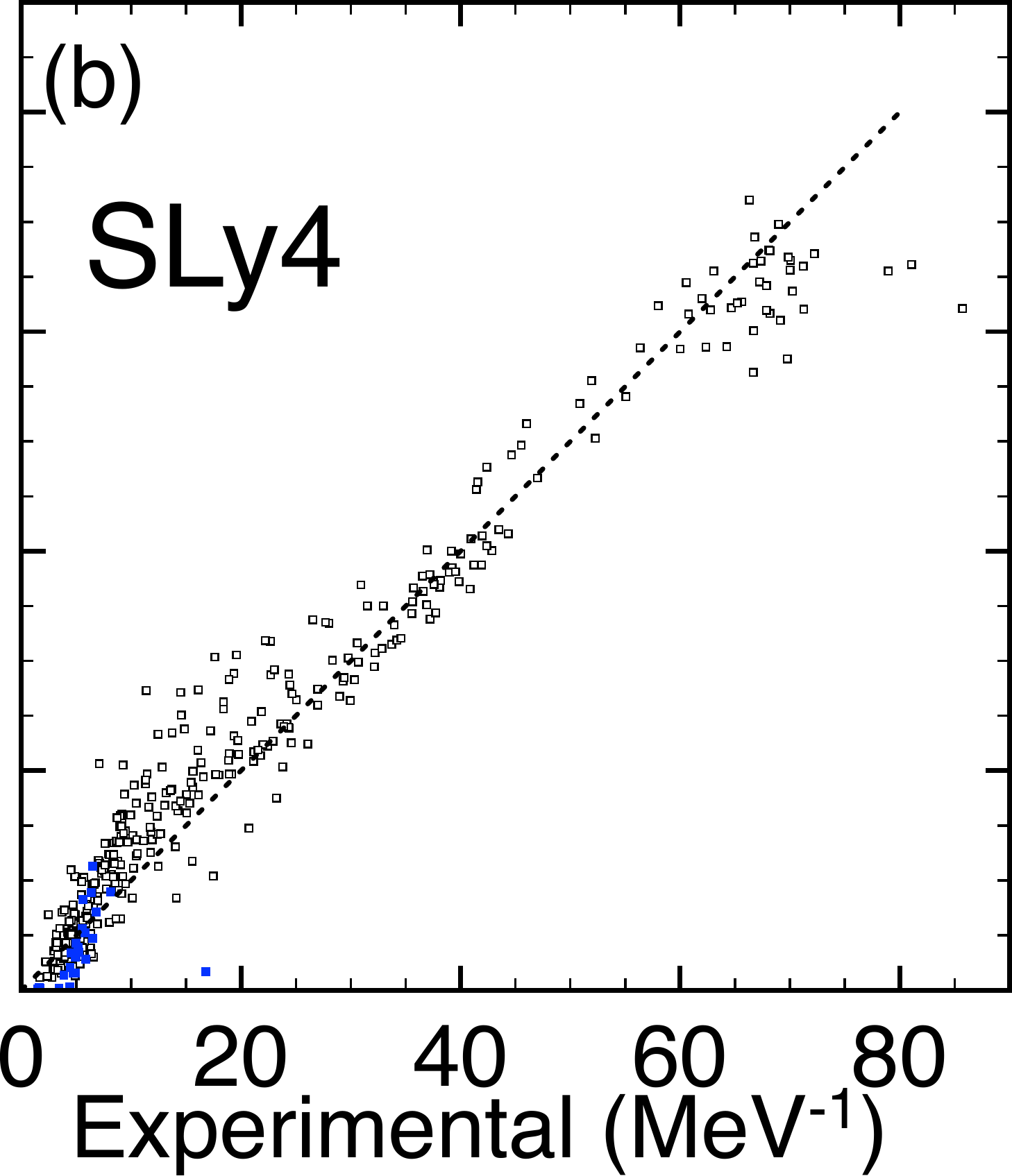}\\ \vspace{0.2cm}
\includegraphics[scale=0.46]{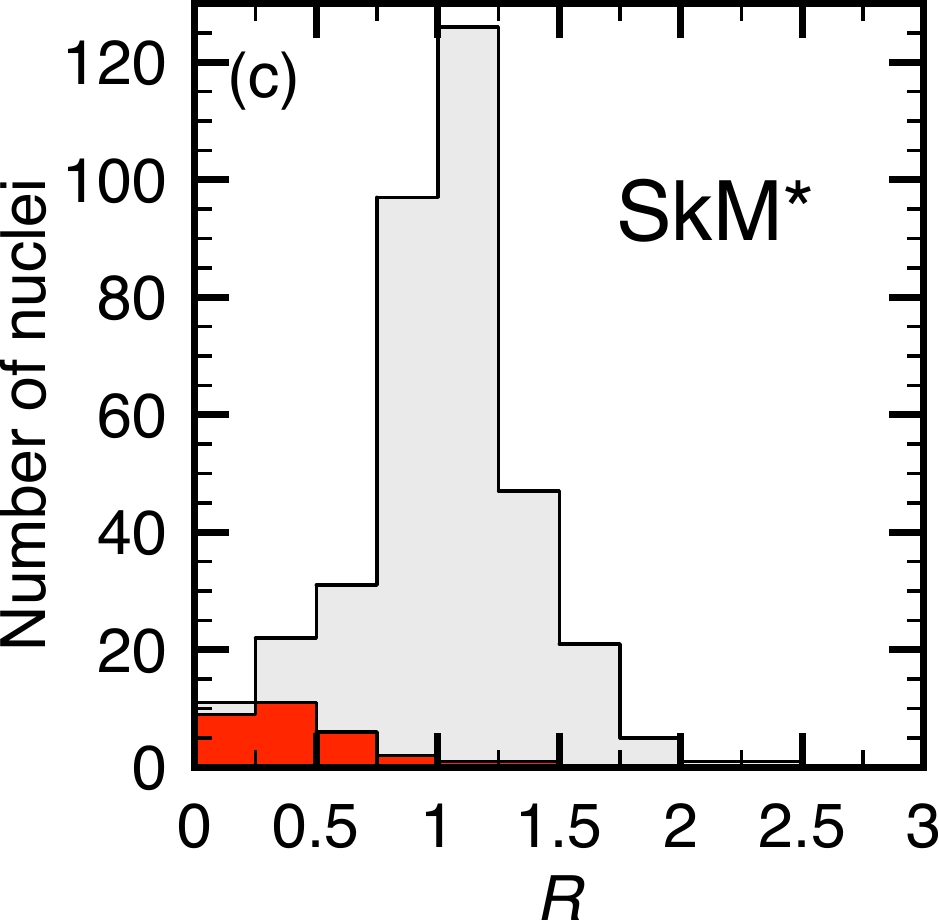}
\includegraphics[scale=0.46]{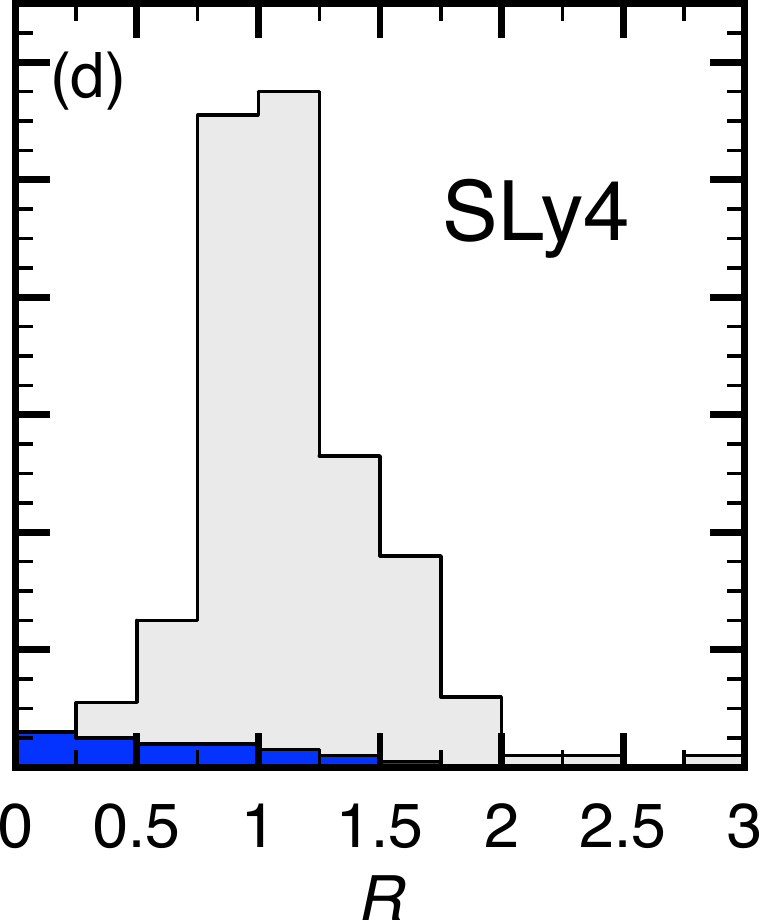}
\caption{\label{fig:MOI_scatt} 
Calculated MoI for 362 nuclei for the SkM* functional (a) and 
375 nuclei for the SLy4 functional (b), plotted versus experimental ones. 
Filled symbols indicate 30 (25) nuclei possessing a weak deformation with $\beta < 0.1$ with 
SkM* (SLy4).
Histogram of the quantity 
$R=\mathcal{J}_{\rm th}/\mathcal{J}_{\exp}$ for the SkM* (c) and SLy4 (d) data set. 
The area in dark indicates nuclei possessing a weak deformation with $\beta < 0.1$.
}
\end{center}
\end{figure}

Figures~\ref{fig:MOI_scatt}(a) and ~\ref{fig:MOI_scatt}(b) 
show the calculated MoI obtained by using SkM* and SLy4
versus experimental ones.
The points follow the diagonal line reasonably well with some scatters that vary in extent over the different regimes.
For transitional nuclei, one may wonder about the validity of the present model. 
The filled symbols in 
Figs.~\ref{fig:MOI_scatt}(a) and \ref{fig:MOI_scatt}(b) 
denote the weakly deformed nuclei having $\beta < 0.1$. 
These nuclei give a small value for the MoI, 
corresponding to higher $E(2^+_1)$ than measurement.
Furthermore, one sees a distinct deviation from the straight line for the highest region around $\mathcal{J}=60$ MeV$^{-1}$: 
$^{238,240}$Cm and $^{244}$Cf.

To make a quantitative measure of the theoretical accuracy, 
I compare theory and experiment, and examine the statistical properties of the quantity $R=\mathcal{J}_{\rm th}/\mathcal{J}_{\exp}$. 
Here $\mathcal{J}_{\rm th}$ and $\mathcal{J}_{\rm exp}$ are the theoretical and experimental MoI. 
A histogram of the distribution of $R$ is shown in 
Figs.~\ref{fig:MOI_scatt}(c) and \ref{fig:MOI_scatt}(d).
For SkM* (SLy4), the average is $\bar{R}=1.02$ (1.16).
When excluding the weakly deformed nuclei with $\beta<0.1$, $\bar{R}=1.07$ (1.15) for 332 (350) data. 
Therefore, the present model overestimates the MoI by about 10\%.

The width of the distribution is an important quantity to determine the accuracy and reliability of the theory. 
One sees that the error is systematic, and the overall distribution is strongly peaked when excluding the weakly deformed nuclei 
that cause a tail in small $R$. 
The root-mean-square deviation, the dispersion, of $R$ about its mean is $\sigma = 0.09$ (0.12). 
Thus, a typical error is about 10\%. 

It is interesting to compare the present calculation with the beyond-mean-field type calculations~\cite{sab07,ber07}. 
The excited $2^+$ states were obtained by the minimization after projection (MAP) and the 
generator coordinate method (GCM) using the SLy4 functional~\cite{sab07} 
or the 5-dimensional collective Hamiltonian (5DCH) 
based on the GCM together with the Gaussian overlap approximation 
using the Gogny D1S functional~\cite{ber07}. 
The authors in Refs.~\cite{sab07,ber07} introduced the measure $R_E=\ln (E_{\rm th}(2^+)/E_{\rm exp}(2^+))$ to evaluate the validity of the theoretical framework. 
Then, I evaluate $E(2^+)$ as $3/\mathcal{J}$ in the present model.

\begin{table}[t]
\caption{\label{tab:comp} 
Statistics for the performance of the CHFB calculations. 
Averages $\bar{R}_E$ and standard deviations $\sigma_E$ for measured $E(2^+)$ are summarized. 
The values for MAP and GCM are taken from Ref.~\cite{sab07}, while 5DCH from Ref.~\cite{ber07}.}
\begin{ruledtabular}
\begin{tabular}{lccc}
model & \# of nuclei & $\bar{R}_E$ & $\sigma_E$ \\
\hline
CHFB (SkM*+YSN) & 332 & $-0.021$ & $0.11$ \\
CHFB (SkM*+mixed) & 325 & $0.029$ & $0.12$ \\
CHFB (SLy4+YSN) & 350 & $-0.095$ & $0.09$ \\
CHFB (SLy4+mixed) & 356 & $-0.053$ & $0.14$ \\
MAP (SLy4) & 359 & $0.28$ & $0.49$ \\
MAP (SLy4, deformed) & 135 & $0.20$ & $0.30$ \\
GCM (SLy4) & 359 & $0.51$ & $0.38$ \\
GCM (SLy4, deformed) & 135 & $0.27$ & $0.33$ \\
5DCH (D1S) & 519 & $0.12$ & $0.33$ \\
5DCH (D1S, deformed) & 146 & $-0.05$ & $0.19$ \\
\end{tabular}
\end{ruledtabular}
\end{table}

Table~\ref{tab:comp} summarizes the statistics for the performance. 
The present model gives 
a compatible description for the average of the energy 
to the 5DCH approach for deformed nuclei. 
The dispersion is better than in other models. 
This comparison indicates that the $2^+_1$ state 
is mostly governed by the rotational MoI of the ground state, 
and the self-consistent cranking model 
describes the $2^+_1$ state surprisingly well for deformed 
nuclei with $\beta > 0.1$.
However, it does not mean the rotational band with the excitation energy $\propto I(I+1)$ always appears in deformed nuclei with $\beta > 0.1$ 
because the MoI in the cranking model depends on spins.

I briefly mention the performance for the intrinsic quadrupole deformation.
For selected nuclei of the Nd and Sm isotopes,
it was demonstrated that 
the mean-field approximation describes well 
the evolution of deformation; see Fig.1 of Ref.~\cite{yos11}.
There are 396 even-even nuclei with known $\beta$~\cite{NNDC}, where 
the deformation parameter is evaluated from the $E2$ transition probability: 
$\beta=(4\pi/3Z R_0^2)\sqrt{B(E2)/e^2}$~\cite{rin80}.
I exclude 10 very light nuclei ($Z<10$). 
An additional 156 (146) spherical nuclei 
have been eliminated as in the above analysis, leaving 230 (240) nuclei.
I define the measure $R_\beta=\ln(\beta_{\rm cal}/\beta_{\rm exp})$ similarly to $E(2^+_1)$. 
I then find $\bar{R}_\beta=-0.12 (-0.11)$ 
with the dispersion $\sigma=0.12 (0.09)$ for SkM* (SLy4), and 
$\bar{R}_\beta=-0.08 (-0.09)$, $\sigma=0.07 (0.05)$ for 219 (233) nuclei with $\beta > 0.1$.
The performance is as good as for the MoI.

\begin{figure}[t]
\begin{center}
\includegraphics[scale=0.35]{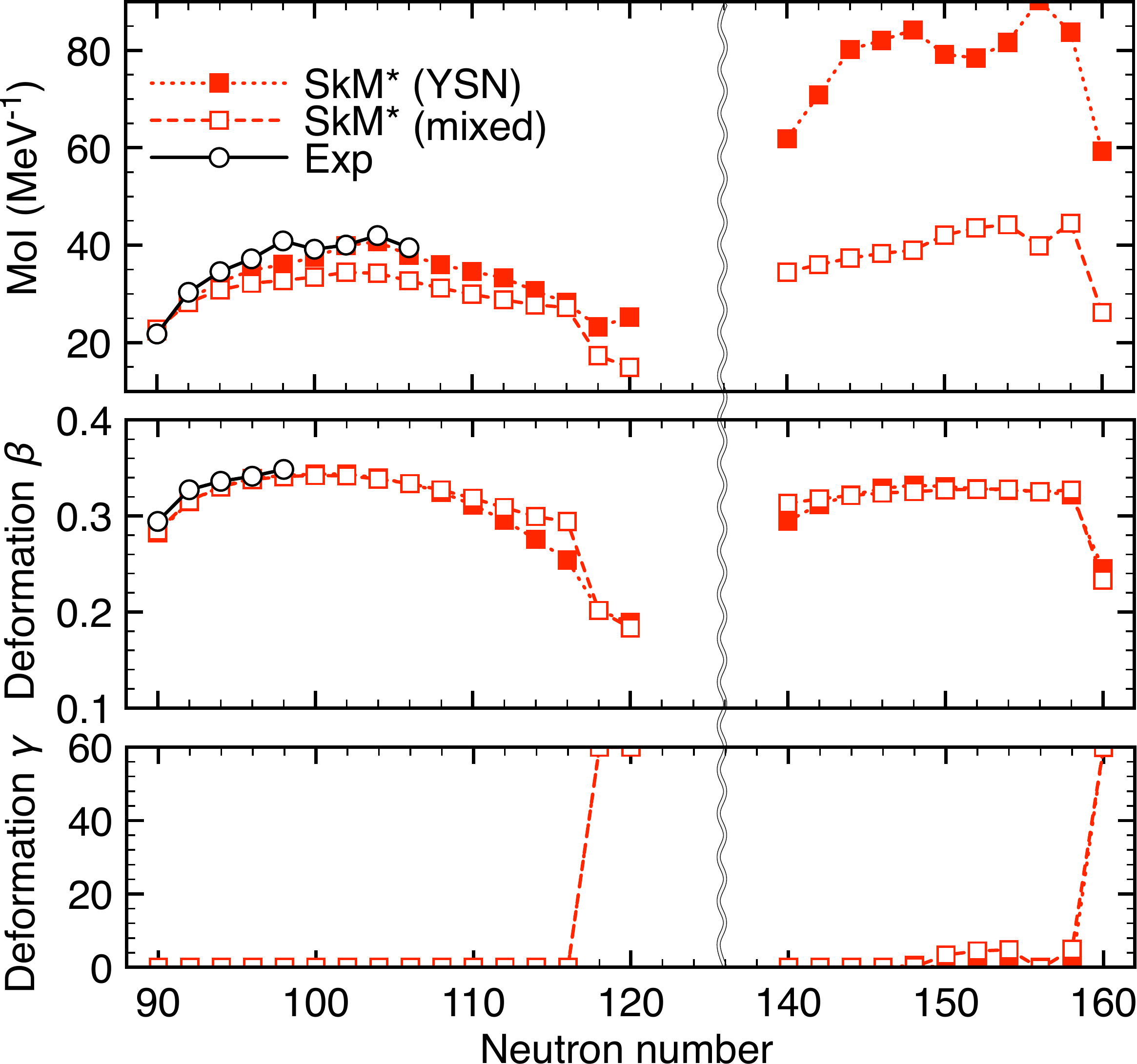}
\caption{\label{fig:Dy} 
(Upper) MoI for the Dy isotopes with 
$N=90\text{--}120$ and $N=140\text{--}160$. 
(Middle) Deformation parameters $\beta$ of protons.
(Lower) Deformation parameters $\gamma$ (in degree) of protons. 
Calculated values are compared with available experimental data~\cite{NNDC}. 
The experimental data for $N=106$ are taken from Ref.~\cite{wat16}.
}
\end{center}
\end{figure}

\subsection{Moments of inertia of drip-line nuclei}
Then, I investigate the MoI of neutron-rich nuclei, 
and discuss unique features near the drip line. 
Figures~\ref{fig:def_MoI}(a) and ~\ref{fig:def_MoI}(c) 
show the calculated MoI. 
I include the even-even nuclei up to $Z=120$ 
and below the magic number of $N=184$.
A striking feature observed in the result shown in Fig.~\ref{fig:def_MoI} 
is that the deformation is strong in the neutron-rich lanthanide nuclei  
around $N=100$ and that the MoI are large accordingly. 
Furthermore, 
the MoI of the rare-earth nuclei near the drip line 
are comparable to those of the heavy actinide nuclei, 
although the mass number is different by about 40.

I take neutron-rich Dy isotopes as an example of rare-earth nuclei, 
and investigate in detail the mechanism of the enhanced MoI near 
the drip line. 
Figure~\ref{fig:Dy} shows the calculated MoI together with
the deformation parameters for $N=90\text{--}120$ and $N=140\text{--}160$. 
These isotopes are well deformed $\beta \gtrsim 0.2$ and 
the estimation of $E(2^+_1)$ is reliable. 
The experimental data for $\beta$ are available up to $N=98$, and 
the present calculation well reproduces the isotopic dependence. 
The $E(2_1^+)$ value is measured up to $N=106$~\cite{wat16}.
Despite the largest deformation being expected at $N=100$, 
the MoI is the largest at $N=98$ and 104. 
The calculation also produces the largest MoI at $N=104$.
This is due to the weakening of the pairing of neutrons: 
The pairing gap energy of neutrons is the lowest at $N=104$ among 
the isotopes with $N=90\text{--}112$. 
The increase at $N=120$ is due to the vanishing pairing gap of neutrons. 
The value of MoI is sensitively determined by the shell effect 
and the pairing rather than the magnitude of deformation.

An exotic behavior shows up when approaching the drip line.
The MoI in the isotopes with $N \sim 150$ 
is about twice as large as that in the $N\sim 100$ region, 
although the deformation of protons is almost the same. 
Since the neutrons are spatially extended, $\beta$ of neutrons and of matter 
are both smaller than those in the $N\sim 100$ region, 
which is against a na\"ive perspective for large MoI.
The pairing is a possible origin of this unique feature near the drip line.

In asymmetric systems, the isovector densities appear to play a role. 
To see the effects of the isovector densities 
in the pairing density functional,  
I perform the calculation without the $\rho_1$ terms in Eq.~(\ref{YSN_pairing}); 
the parameters $\eta_1, \eta_2$ are set to zero while keeping $\eta_0=1/2$. 
This corresponds to the mixed volume and surface. 
The strength $V_0$ in Eq.~(\ref{pair_DF}) was fixed to the pairing gaps of $^{156}$Dy 
for the YSN functional~\cite{yam09}. 
I found the strength 
$V_0=-289$ ($-326$) MeV fm$^3$ and $-324$ ($-343$) MeV fm$^3$ 
for neutrons and protons with SkM* (SLy4) produces the same pairing gaps 
to the ones obtained using the YSN functional. 
The performance for describing $E(2^+)$ is as good as 
the YSN functional, as listed in Table~\ref{tab:comp}. 
The calculated MoI are displayed in Figs.~\ref{fig:def_MoI}(b) and ~\ref{fig:def_MoI}(d).

The results for the neutron-rich Dy isotopes are shown in Fig.~\ref{fig:Dy}. 
The magnitude of deformation is similar to that calculated with the YSN functional 
for both in the $N\sim 100$ and $N \sim 150$ regions.
The calculated MoI are slightly smaller around $N=100$, 
and a 16\% reduction is found at $N=104$. 
A deformed-shell effect sensitively affects the MoI when 
using the YSN functional.
With 
the mixed-type pairing, 
the pairing gap of neutrons decreases gradually from $N=90$ to 114.
Near the drip line, the mixed-type pairing gives 
almost 50\% reduction of MoI to the YSN pairing.

\begin{figure}[t]
\begin{center}
\includegraphics[scale=0.25]{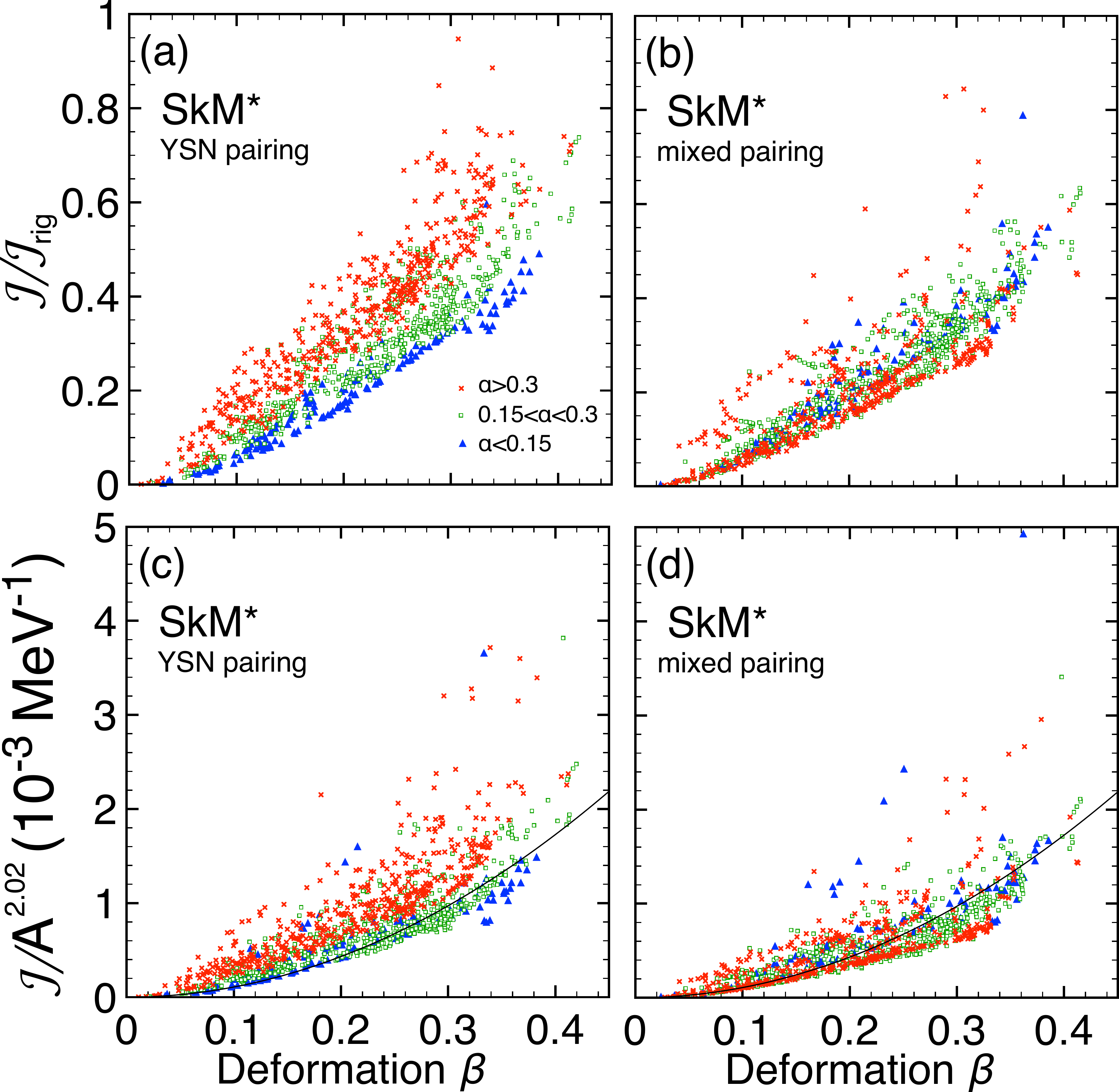}
\caption{\label{fig:MoI_ratio} 
Ratio of the MoI calculated 
by using the YSN (a) and mixed-type (b) pairing functional 
and the ones for the rigid body. 
The calculated MoI devided by $A^{2.02}$ as a function of the 
deformation parameter $\beta$ using the YSN (c) and mixed-type (d) 
pairing. 
The line represents Eq.~(\ref{eq:emp_J}) 
with the approximation $R_0=1.2A^{1/3}$ fm. 
Symbols with cross, square, and triangle indicate the asymmetry parameter $\alpha \ge 0.3$, 
$0.15 \le \alpha<0.3$, and $\alpha < 0.15$, respectively.
}
\end{center}
\end{figure}

A significant enhancement of MoI near the drip line 
using the YSN functional 
is due to  
a deformed-shell effect and 
an isovector-density dependence (an effective decrease in the strength) 
of the pairing functional. 
Indeed, this mechanism explains the lowering of $E(2_1^+)$ in $^{40}$Mg~\cite{yos22}.
The reduction in the strength of the pair interaction with 
an increase in the asymmetry 
can be seen by the comparison of Figs.~\ref{fig:MoI_ratio}(a) and \ref{fig:MoI_ratio}(b).
A reduction of the MoI relative to the rigid body is due to the pairing, 
and the reduction found 
in very neutron-rich nuclei with the asymmetry $\alpha=(N-Z)/A >0.3$ is 
apparently weakened when using the YSN pairing functional. 
Scattering of the data points is associated with the shell effect. 

Finally, I discuss the enhancement of MoI in a different point of view. 
As the quadrupole collectivity increases, one sees 
a lower energy and a stronger transition. 
Empirically, the following relation has been found 
and the 91\% of the observed 328 data points are reproduced 
within a factor of two~\cite{ram01}: 
\begin{equation}
    \left[\dfrac{B(E2;0^+_1 \to 2_1^+)}{1\, e^2 \rm{fm}^4}\right]\times 
    \left[\dfrac{E(2^+_1)}{1\, \rm{MeV}}\right]=32.6 \dfrac{Z^2}{A^{0.69}}.
\end{equation}
This corresponds to
\begin{equation}
    \mathcal{J}=\dfrac{3}{32.6}\left(\dfrac{3}{4\pi}\right)^2 
    A^{0.69}R_0^4 \beta^2
    \hspace{0.2cm}[{\rm MeV}^{-1}],
    \label{eq:emp_J}
\end{equation}
where $R_0$ is given in the unit of fm. 
Figures~\ref{fig:MoI_ratio}(c) and \ref{fig:MoI_ratio}(d) show 
the calculated MoI divided by $A^{2.02}$ as a function of 
the deformation parameter $\beta$.
With the mixed-type pairing, 
the calculated MoI scatter around the empirical line, 
and most of them are within a factor of two.
On the other hand, with the YSN functional, 
the empirical line is entirely 
off the trend of the calculated MoI for $\alpha>0.3$.
Therefore, $E(2^+_1)$ can be low in neutron-rich nuclei
despite the $B(E2)$ value is not high. 
A systematic measurement of $E(2_1^+)$ and $B(E2)$ in 
neutron-rich nuclei deepens the understanding 
of the pairing in nuclei 
and puts a constraint on the pairing density functional.

\section{Summary}
I have performed systematic calculations of the MoI 
from the proton drip line to the neutron drip line 
to see the roles of neutron excess 
in the collective rotational motion. 
To describe neutron-rich nuclei where 
the loosely-bound neutrons and 
the continuum coupling are necessary to consider, 
the cranked HFB equation is solved 
in the coordinate space.
The comparison with the available experimental data and 
other models shows that 
the present model surprisingly well describes the 
ground-state MoI, namely the $E(2_1^+)$ value, 
for deformed nuclei with $\beta > 0.1$. 
By employing the pairing density functional constructed 
to describe the isospin dependence in neutron-rich nuclei, 
I have found that the MoI are greatly enhanced near the drip line, 
whereas the magnitude of deformation is not as strong as estimated 
by the empirical relation between the $E(2^+_1)$ and $B(E2)$ values.
A systematic measurement of $E(2^+_1)$ and $B(E2)$ 
in neutron-rich nuclei 
puts a constraint on the density dependence of 
the pairing effective interaction. 
The stronger the isovector-density dependence is, 
the more significant the enhancement of MoI in neutron-rich nuclei is.

\begin{acknowledgments} 
This work was supported by the JSPS KAKENHI (Grants No. JP19K03824 and No. JP19K03872). 
The numerical calculations were performed on the computing facilities  
at the Yukawa Institute for Theoretical Physics, Kyoto University, 
at the Research Center for Nuclear Physics, Osaka University, and 
at the Cybermedia Center, Osaka University.

\end{acknowledgments}

\end{document}